\documentclass[twocolumn]{article}
\usepackage{authblk}
\usepackage{graphicx}
\usepackage{amsmath}

\usepackage[utf8]{inputenc}
\usepackage{graphicx}
\usepackage{xcolor}

\newcommand{\jb}[1]{\textcolor{black}{#1}}  
\newcommand{\mc}[1]{\textcolor{black}{#1}}     
\newcommand{\dc}[1]{\textcolor{black}{#1}}   

\begin{document}


\title{Softer than soft: diving into squishy granular matter}

\author[1]{Jonathan Bar\'{e}s}
\author[1]{Manuel C\'{a}rdenas-Barrantes}
\author[2]{David Cantor}
\author[1]{Mathieu Renouf}
\author[1,3]{Émilien Az\'{e}ma}
\affil[1]{LMGC, Universit\'{e} de Montpellier, CNRS, Montpellier, France.}
\affil[2]{Department of Civil, Geological and Mining Engineering, Polytechnique Montr\'{e}al, Montr\'{e}al, QC, Canada.}
\affil[3]{Institut Universitaire de France (IUF), Paris, France.}


\date{\today}

\maketitle

\abstract{
Softer than \textit{soft}, \textit{squishy} granular matter is composed of grains capable of \dc{significantly changing} their shape \jb{(\dc{typically} larger than $10$\% \dc{of deformation})} without \dc{tearing or} breaking. Because of the difficulty to test \dc{these materials} experimentally and numerically, \dc{such a} family of \dc{discrete} systems remains largely ignored in the granular matter physics field \dc{despite being commonly found in nature and industry}. Either from a numerical, experimental, or analytical point of view, the study of highly deformable granular matter involves several challenges covering, for instance\dc{:} ($i$) the need to include a large diversity of grain rheology, ($ii$) the need to consider \dc{large material} deformations, and ($iii$) the analysis upon the effects the large \dc{body distortion has on} the global scale. In this article, we propose a thorough definition of these \dc{squishy} granular systems, and we summarize the \dc{upcoming} challenges in their study.}

\section{Introduction}

\begin{figure}
\centering
\includegraphics[width=\linewidth]{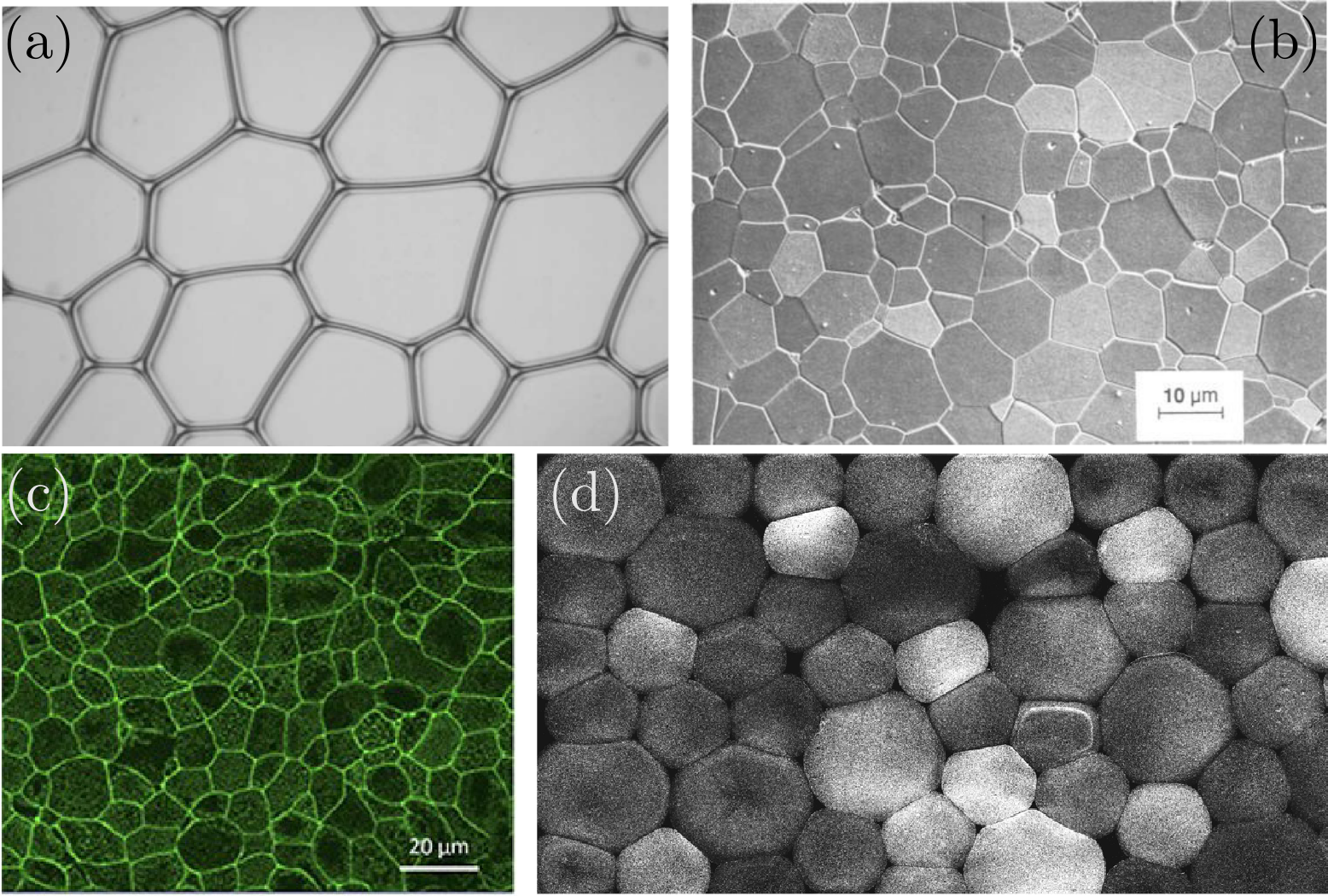}
\caption{Squishy granular systems. (a) Highly deformed 2D soap bubbles from \cite{kabla07_jfm}. (b)  Sintered ceramic from \cite{rhodes95_bk}. (c) Jammed layer of biologic cells from \cite{osullivan2020_ob}. (d) Plastic particles made of agar gel deep in the jammed state from \cite{vu2019_pre_0}. }
\label{fig_1}
\end{figure}

The importance of granular matter for human activities is quietly striking. It not only covers $90$\% of the Earth's crust \cite{Nataf00_bk}, it \dc{adds up to} about $70$\% of the materials used in the industry, represents the second-largest volume of matter carried by human beings -- just behind water \cite{lechenault07_phd} --, and more energy is used to produce it than that consumed in human transportation \cite{guyon20_bk}. For all these reasons, a thorough understanding of the behavior of granular matter made of all sorts of materials is of tremendous importance for \dc{humankind}, from environmental preservation \dc{and} control of geohazards to space exploration. Yet, this is an extremely complex task since these materials span over several scales, from \dc{colloids to screes or asteroids}. These materials can also be quite different -- made of virtually any sort of matter -- they can behave like a solid, a liquid, or a gas depending on loading conditions and local interactions. 
Over the last three decades, a framework and state equations to describe the flow of granular systems made of rigid\jb{, virtually undeformable,} particles have been set up \cite{midi04_epje,jop06_nat,coquand21_pre}. Similarly, the jamming transition of \jb{these rigid grains has been fruitfully explored and the so obtained results have been extended to the case of soft grains} \cite{liu98_nat,ohern03_pre,bi11_nat,zhao19_prl}. Nevertheless, \dc{it is} unsettling that in much of the literature involving deformable granular media the material these particles are made of is assumed to stay in the linear elastic regime so that the Hertz contact assumption remains valid \cite{archard57_prsl,johnson85_bk}. \jb{This means that the level of this material deformation does not exceed $10$\%}. 
\dc{However, we can} naturally wonder about the consequences of such an assumption \dc{in} granular systems undergoing more intense loading, \jb{well above $10$\% of deformation \cite{vu2019_em}}. The existing framework mostly neglects the grains' change in shape, limiting the number of real-life scenarios that can be described: \jb{This approach is good at describing the loading of corn grains but not of pop-corn.}As shown in Fig. \ref{fig_1}, many systems, ranging from biology to industrial applications, are made of particles brought to large deformation regimes. To differentiate these systems from those composed of \textit{soft} grains, we propose to call \textit{squishy} granular materials \dc{all discrete matter whose individual bodies undergo large nonlinear deformations}.
This definition implies that the global stresses undergone by the system are substantially higher than the Young modulus of the particles, \textit{i.e.}, $E_{\rm{material}} \ll \sigma_{\rm{loading}}$.

\dc{In addition}, stating that a granular system is \textit{squishy} \dc{also} means that the particles' change of shape must be considered to describe their local interactions, displacement, and, more generally, to understand the global behavior of the granular medium.
In other words, the exact nature of a squishy material cannot be hidden behind a linear elastic law. The intrinsic rheology of the particles becomes paramount and must be considered through the nonlinear theory of elasticity \cite{taber2004_bk}. One step further, the exact geometry of the particles must be thoroughly considered since it can, for example, induce discontinuous and sudden phenomena such as buckling. This is the case for shell particles \cite{jose15_sm,boromand2018_prl} or, more generally, for particles made of any metamaterial \cite{florijn14_prl,reis15_eml}. 
Naturally, a large diversity of geometries and material behaviors have to be considered to properly handle squishy media. This can encompass thermal and anisotropic effects, as well as magnetic interactions \cite{hernandez17_epj,cox16_epl}.  
As a limitation in the broad category of squishy particles, we assume that the number of entities in the system stays stable all along the loading process. The squeezable nature of the particles allows them to support any loading with no merging or fragmentation.
Thus, the edge of each particle is continuously described by a closed function.

Fortunately, the granular matter physics field is not starting from zero as far as it concerns squishy particles. The most advanced studies in this domain come from the rheology of liquid foams \cite{hohler05_jpcm}, materials composed of gas bubbles within liquids. Numerous experimental and numerical works have \dc{significantly contributed to the understanding} of these materials under different loadings \cite{debregeas2001_prl,asipauskas2003_gm,clancy2006_epje,kabla07_jfm} and direct links have been drawn to granular matter physics \cite{weaire07_bk}. However, emulsions and colloids wildly differ from other discrete materials \dc{when considering} their material rheology and inter-particle frictional properties \cite{brujic03_fd,newhall12_prl,krebs13_sm,hohler19_acis}.
Consequently, it is difficult to generalize the results found for foams, colloids, and emulsions to the case of squishy granular matter. 
\dc{Motivated} by technological applications, very soft grains have also been studied under compression as in the case of sintered material manufacturing. The compaction of ceramic, metal, gels, and drug powders has been \dc{central in the research of} compacted materials \cite{cooper1962_jacs,kawakita71_pt,wu05_pt} or self-formed colloids \cite{fan17_sml}. In \dc{all} these cases, \dc{however}, studies are \dc{still} limited to the plastic behavior of particles under compression. 
Other examples of squishy grains can be found in living organisms. Cells forming a tissue or clogging in a vein \dc{can certainly be described using squishy particle assemblies} \cite{newhall12_prl,park15_nat,wyatt15_pnas,mauer18_prl,osullivan2020_ob}. 

From a numerical point of view, only recently \dc{a number of} models have emerged for squishy granular matter. Some of them, based on the discrete element method (DEM), \dc{have allowed researchers to catch a glimpse of the behavior of multiparticle, highly deformable systems} \cite{nezamabadi2017_gm,boromand2018_prl,wang2021_sm}, \dc{but} remain unsatisfactory in terms of material mimicking. Other methods based on finite-elements (FEM) are better at mimicking particles' materials, but have difficulties to properly take contact interactions into account \cite{mollon2018_gm,vu2019_pre,cantor2020_prl,cardenas2022_sm}. In both cases, major improvements are necessary to simulate any squishy granular system.   

The rest of this article is dedicated to summarizing the different loading conditions that squishy granular matter can undergo and the different approaches to deal with their characterization, simulation, and analysis. More particularly, we focus on the experimental and numerical challenges to understand the behavior of squishy grains under compression and shearing.

\section{Browsing deep in the jammed state}
\subsection{Compressing squishy particle assemblies}

The mechanical behavior of a granular system is ruled by the individual behavior of each particle within it. For example, when hard particles -- \textit{i.e.,} made of a material with high stiffness relatively to the applied stress -- are compressed inside a box, they tend to rearrange compactly. In this case, the rearrangement ends when a stable force network is reached. This state, called the jammed state, depends mainly on the morphology of the particles, their size and shape \cite{vanhecke09_jpcm}, and their interaction laws \cite{Torquato2010_Jammed,vanhecke09_jpcm,Silbert2010_jamming}. If one considers assemblies of soft particles with low stiffness \jb{relatively} to the applied stress, they \dc{are able to} undergo large deformations, and the compression can continue beyond the jammed state. 
Exploring the compaction of granular matter and, especially, systems very deep in the jammed state is challenging, mainly due to the nature of the particle rearrangements and the evolution of the force-chain network. 
For assemblies of squishy particles, the compactness -- \textit{i.e.,} the ratio between the volume of the particles and the volume of the box -- surpasses the compactness of the jammed state\jb{, the random close packing,} and tends to a maximal value close to $1$. 
Naturally, this high compactness comes with an increasing number of contacts and contact surfaces. Therefore, it is seen as an evolution of the probability distribution of the forces and an evident reorganization of the force-chain network that drastically change the mechanical properties of the assembly.

Innovative experiments and advanced numerical methods have allowed one to take a small step forward in understanding the microstructural evolution beyond the jamming point (see Section {\bf \ref{Exp_Num}}). However, theoretical modeling of the compaction process remains \jb{a complicated} task. Although many equations \dc{have been proposed in the last decades to} try to link the confining pressure to the compactness  \cite{Heckel1961,Kim1987,Secondi2002_Modelling,Cabiscol2020_Effect}, most of them are based on empirical strategies with no apparent physical bases, \dc{requiring} several fitting parameters. Only recently, a systematic approach, based entirely on micromechanical parameters and free of any {\it ad hoc} parameters, has been proposed and numerically tested in the case of hyperelastic circular, polygonal, and spherical shapes under different frictionnal conditions \cite{cantor2020_prl,cardenas2022_sm}. This method was derived from the micromechanical expression of the granular stress tensor together with the approximation of the Hertz contact law between two particles at small strain and the theoretical solution of the hydro-static compaction of a single particle. 

Although the above micromechanical approach \dc{constitutes a substantial progress} in the challenge of understanding and modeling the compaction of deformable particles, it only describes convex particles under isotropic compaction with continuous laws of elasticity. Many unknowns remain unresolved concerning particles with a more complex shape or composition characteristics, such as non-convex particles or particles that do not follow elastic or continuous deformation behavior \dc{(for example, particles suffering buckling)}. Also, we still need to better understand the behavior of granular systems under different loading conditions, such as uniaxial compaction or shear loading.

\subsection{Shearing squishy particle assemblies}

The second and more complex loading that a squishy granular system can undergo is shearing. This loading type induces not only large particle deformations but also large and often erratic particle rearrangements. When this latter regime dominates, the granular system yields and flows. Below the jammed state, for rigid grains, the flowing regime is described in many configurations by well-known state equations \cite{midi04_epje,jop06_nat,coquand21_pre}. 
Although squishy grains \dc{could} present state equations reminiscent of those for rigid grains, this has not been shown \jb{yet}. 
So far, for a dense flow of soft grains, it has been shown that grains rearrange through a succession of local mechanisms called T1-events \cite{kabla07_jfm,bi2014_sm} inducing specific glassy rheology. These T1-events, \jb{presented in Fig.\ref{fig_2}(e)}, mainly happen in a relatively narrow area of the system called \jb{the} shear band. This has been observed for liquid foams \cite{kabla07_jfm} and living cells \cite{bi2014_sm}, but one can still wonder how the shear-band evolves depending on the material rheology. \dc{Do} these T1-events still exist for other material types such as plastic ones?

The shearing of granular systems often \dc{exhibits} erratic dynamic mechanisms at the global scale -- also known as avalanches -- even \dc{under} constant loading \cite{daniels2008_jgrse,bares2017_pre}. For soft grains, depending on the particle stiffness, strain rate, and confining pressure, the global dynamics can display different statistical signatures \cite{zadeh2019_pre}. In densely packed squishy systems, the space for particles to rearrange can be extremely reduced and the material rheology has a strong influence on the possible rearrangement mechanisms. As such, it is expected that the avalanche dynamics are strongly modified by the exact particle nature and frictional properties. But we still do not know how or up to which extent this is the case.

In the case of rigid or soft particles, plasticity, seen as an irreversible deformation of the material, happens at the meso-scale: grains rearrange so \jb{that} they cannot recover their previous position when the loading is released. In the case of squishy granular systems not only this plastic mechanism is at play, but also the intrinsic plasticity of the particles. Hence, the plasticity of these amorphous materials is by definition of a different nature as the one of soft grains.

Concerning the shearing of squishy particles, the main challenge is to evidence the local rearrangement mechanisms, their dynamics and spatial distribution, and to understand how they induce the flowing of the sheared system. The second challenge consists in understanding the coupling between meso and micro-scale plasticity processes in squishy amorphous materials 

\section{Experimental \& Numerical challenges}
\label{Exp_Num}
\subsection{Physical experiments}

Studying experimentally granular matter is challenging because \dc{its multi-scale nature, so} tests have \dc{to be able} to \jb{catch main features} \dc{across scales}. One needs to follow the behavior of individual grains that \dc{end up inducing} \jb{the global material behavior from specific particle properties}. In the case of squishy granular matter, one also needs to go one scale \jb{smaller} since it is necessary to consider the material behavior inside each grain at the \textit{sub-granular scale} \cite{vu2019_pre_0}. Although several studies are possible without analyzing this sub-granular \dc{level} or measuring the local stresses -- as it has been done in the early stages of the granular matter physics -- a comprehensive study of squishy granular matter \jb{requires} such a multiscale approach. 

So far, whether it is in a 2D or 3D configuration, the existing ways to measure grain displacements, contacts, and stresses are all based on an inverse problem method. Usually, the Hertz contact law \cite{archard57_prsl,johnson85_bk}, or a certain surface tension in the case of liquid foams \cite{kabla07_jfm}, is assumed and some local field measurements permit to detect contacts and guess interaction forces. These local measurements can be either the grain boundary deformation \cite{brujic03_fd,kabla07_jfm,brodu2015_natcom}, a photoelastic \cite{daniels17_rsi,zadeh19_gm} or thermoelastic \cite{jongchansitto14_sm} signal, or a digital image correlation (DIC) rough measurements \cite{hurley2014_jmps}. By definition, these measurements are limited by the method sensitivity and the domain of validity of the assumed contact law. \jb{For example in the case of photoelasticity, when deformation is too large, interference fringes overlap and the measured signal saturates.}
Assuming that there is no saturation of the measured signal for large deformations, it would be possible to extend these methods by finding contact laws that include complex material rheologies. For example, if the Hertz contact law could be extended beyond the case of a linear elastic behavior with a finite friction coefficient \cite{vu2019_em}, this would permit extending these methods to squishy grains. Still, one can imagine that the number of parameters to guess in the problem inversion would be much larger, making it complicated to implement. 

Recently it has been proposed not to make any assumption and to directly measure displacement fields at the grain scale, even in the large deformation regime. This process is based on a large field and an accurate imaging method \cite{vu2019_em,vu2019_pre_0,cardenas2021_arx}\dc{, which allows} allows \dc{one} to measure all the tensors of the non-linear elasticity theory and to guess the stress tensor knowing the loading history and the material rheology via large deformation DIC. However, this only works for small systems ($\sim 100$ grains) under quasistatic conditions.
This method -- and the fact that it provides strain fields inside the grains -- comes with a new outlook on this topic upon the notion of force chain and its relevance or validity in the study of densely packed configuration. 

Although recent experimental approaches have permitted to follow the evolution of squashy granular systems, \dc{the development of} an experimental method to characterize squishy particle assemblies under dynamic loading remains a challenging task.

\subsection{Numerical simulations}

\begin{figure}
\centering
\includegraphics[width=\linewidth]{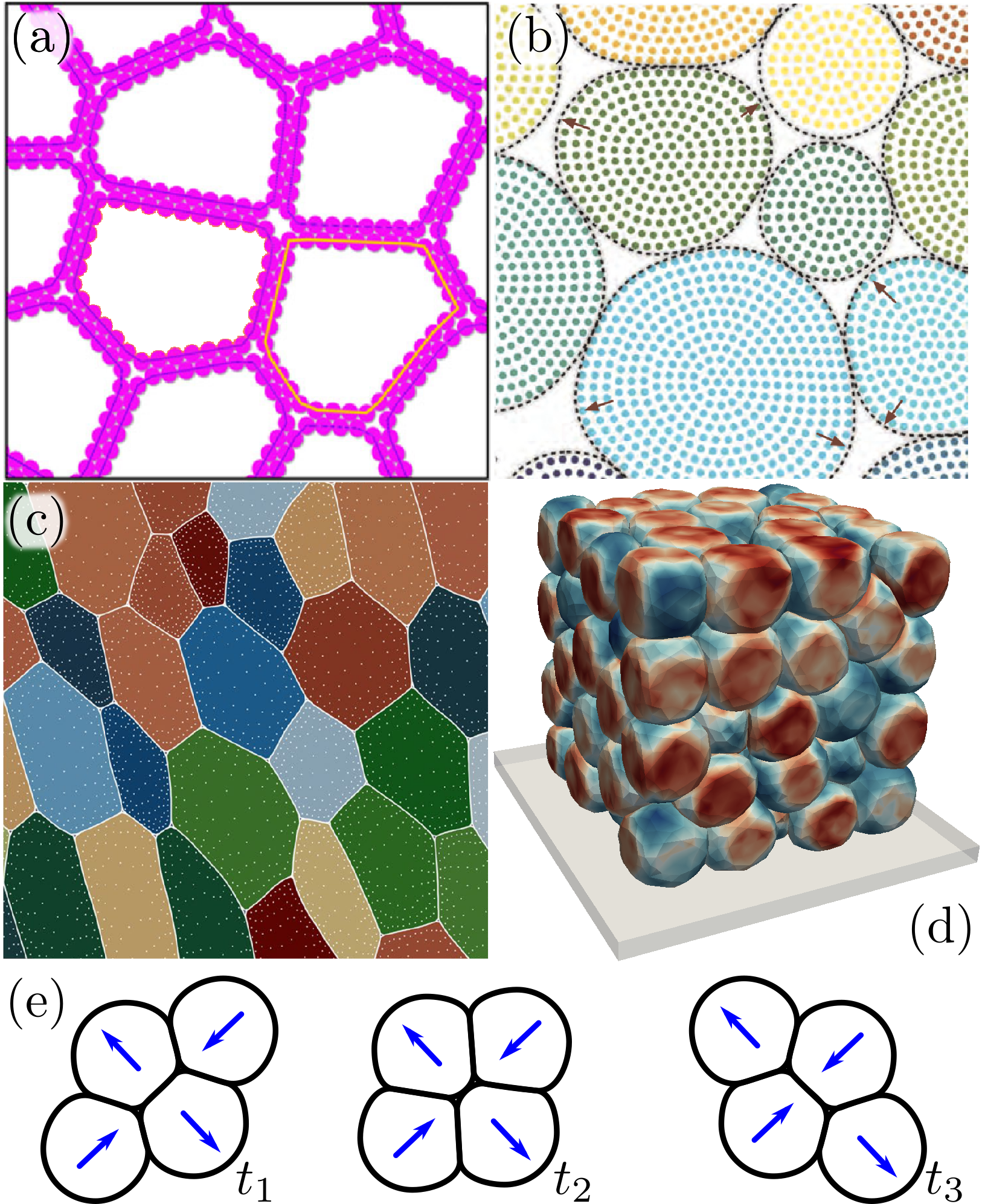}
\caption{(a to d) Modeling squishy particle assembly: (a) using the ``Deformable-Particle" (DP) method, a class of BPM-like method (with permission of C. O'hern et al. \cite{boromand2018_prl}), (b) using the meshless approach MPM (with permission of S. Nezamabadi et al. \cite{nezamabadi2017_gm}), (c) using the meshless ``soft-DEM" approach (with permission of G. Mollon \cite{mollon2022_gm}), (d) using the 3D Non-Smooth Contact Dynamics approach by M. Cardenas-Barrantes et al. \cite{cardenas2022_sm}. \jb{(e) Example of a T1-event running from time $t_1$ to $t_3$. Particles react to shear loading by this peculiar rearrangement: particles in the expansion direction split apart and permit the ones in the compression direction to get closer up to the point where they touch each other.}}
\label{fig_2}
\end{figure}

The modeling of granular media is historically associated with discrete-element approaches (DEM) to describe the dynamical evolution of a collection of hard grains. DEM can be split into two main approaches: the ($i$) smooth and ($ii$) non-smooth \dc{methods}. This classification is related to \jb{the way} a contact between two particles is considered. In the case ($i$), regularised contact laws link the contact force to a slight overlap between particles \jb{in contact}. In the case ($ii$), the unilaterality of contacts (non-regularized laws) traduces the non-penetration of the particles.
Both approaches have proven to produce equivalent results as far as the basic hypotheses of each method are \dc{respected} -- \textit{i.e.,} binary contact and small grain deformations compared to the grain size.

Extending the numerical approaches, first to soft, and further to squishy grains, \jb{implies} new assumptions and new numerical strategies. 
In the framework of \jb{the} smooth approach, \jb{one} \dc{alternative is to increase} the allowed overlapping between bodies \dc{that can be later interpreted as a virtual} grain deformation.
Although this approach is debatable in terms of realistic representation of the grain shapes, it has shown to be valid in the small deformation domain, \dc{as in the case of} the compaction of hard grains assemblies just after reaching the jammed state \cite{Agnolin2008_On,Bonnecaze2010_Micromechanics,Lopera2017_Micromechanical}.
Nevertheless, going deeper into the jammed state leads to \jb{the breaking of some of the model assumptions} and the classical methods need to be extended to meet this new range of deformation. 

As a first approximation, a squishy grain can be considered as an `aggregate' of elements (in the numerical sense) bonded together by bilateral or cohesive bonds. 
The bonding law can be adjusted to mimic the macroscopic behaviour of a given grain material. 
This aggregate representation can be performed throughout the volume of the grain. This approach is known as the Bonded-Particle Method (BPM) \cite{Potyondy2004_A,Utili2008_DEM,Cho2007_A,Asadi2018_Discrete}. Another alternative is to consider sub-elements only on the particle's surface, as proposed in the Deformable Particle Method (DPM) \cite{Chelin2013_Simulation,boromand2018_prl} (see Fig. \ref{fig_2}-a).
The main advantages of these methods are their low computational cost and ease of implementation in the DEM framework. However, in the case of 'aggregate' methods, very large deformations can lead to non-physical situations.
\jb{Another} option consists in using constitutive models to describe the bulk behavior of the particles. Then, it is possible to \jb{use} standard numerical methods such as a finite elements (FEM) or meshless approaches. 

In a meshed approach, a volumetric rheological model associated with a state equation is considered for each grain. In this case, FEM-like methods coupled with DEM are used to describe the particle deformations, keeping -- or not -- its integrity \cite{Procopio2005, Harthong2009_Modeling, Huang2017, Wang2020_Particulate}. 
The advantage of these strategies is to closely \jb{mimic} the geometry and bulk properties of the particles. However, they end up being computationally expensive. 
As with the classical DEM approaches, it is possible to distinguish two classes of methods: the Multi-Particle Finite Element Method (MPFEM) \cite{Gethin2002_A,Xin2003_Investigation, Procopio2005, Frenning2015_Towards}, based on regularized contact interactions, and the Non-Smooth Contact Dynamic Method (NSCD) \cite{Moreau1994_Some,Jean1999}, based on non-regularized contact laws.
MPFEM simulations have been performed by Mesarovic \emph{et al.} \cite{Mesarovic2000_Frictionless} to study the deformation mechanisms in a periodic configuration of dissimilar elastoplastic spheres. For squishy granular packings, the first MPFEM studies were conducted by Gethin et al. in 2002 \cite{Gethin2002_A} and Chen et al. in 2006 \cite{chen2007_Elasto} in 2D and 3D, in the context of metal compaction. The NSCD was used for the first time in 2000 by Acary et al. \cite{Acary1998_Numerical} with a work on masonry. They were followed by Cao et al. in 2011 \cite{Cao2011_Thesis}, who tested the numerical approach for the compaction and shearing of soft packings. The latest works related to compaction of squishy granular packings using NSCD corresponds to Vu et al. \cite{vu2019_pre,Vu2021_Effects} in 2D, followed by Cardenas et al. in 2D and 3D \cite{cardenas2022_sm} (see Fig. \ref{fig_2}-d).

In the framework of mesh-free models, squishy grains have been simulated via the Material Point Method (MPM) \cite{nezamabadi2017_gm} (see Fig. \ref{fig_2}(b)), or the Soft Discrete Element Method (SDEM) \cite{mollon2022_gm} (see Fig.\ref{fig_2}(c)), where simple kinematics are postulated in order to represent the ovalisation of the grains and their local deformations around interparticle contacts.
In the last case, a continuous deformation field is interpolated between the nodes belonging to a grain by means of a moving least square algorithm. These approaches require the calibration of some fundamental computational parameters from experimental data, such as the stiffness and viscosity \cite{Nezamabadi2019_cpc, mollon2022_gm}.
\dc{Despite the development of these alternative methods}, the different results obtained via FEM-DEM \dc{compare better to experiments} and \dc{show} that the behavior of squishy grain collections can only be understood using these advanced numerical approaches despite being extremely time-consuming. 

Most of today's simulations \dc{involving squishy grains} are \dc{performed} in \jb{the quasi-static regime and} under uniaxial deformation. These conditions remain relatively simple and involve few dynamical effects. Other loading conditions such as shearing would rapidly increase the complexity of the simulations, making it difficult to conduct \dc{numerical experiments} at higher shear rates or compression levels. Moreover, while several thousands of squishy grains can be modeled in 2D, only a few hundred can be simulated in 3D in a reasonable computing time. Therefore, \dc{additional} challenges \dc{concern} ($i$) the development of scalable and highly parallel algorithms, ($ii$) the implementation of fully periodic conditions, allowing for the volume changes to account for the expansion/contraction of the particles, \dc{and} ($iii$) \jb{considering} realistic mixtures that include \dc{grains} with different size, shape, and bulk behavior within \mc{the} same system. 

\section{Summary \& Perspectives}
Deep in the jammed state\jb{, when grain deformation is well above $10$\%}, the \textit{soft} granular matter framework is not \jb{relevant} to describe the evolution of granular systems \mc{anymore}. \dc{As} packing fraction grows way beyond the jamming point, it is necessary to consider the large particle change in shape and it effects: ($i$) creation of new contacts, ($ii$) interaction stresses depending on the particle material, and ($iii$) possible large particle rearrangements. To make up for the lack of a correct framework for these systems, we propose the concept of \textit{squishy} granular materials in which \jb{the linear elastic behavior is overpassed to consider} hyperelastic, plastic or other complex material rheologies. 

Considering squishy materials rapidly increases the complexity of the experimental, numerical and theoretical approaches \dc{needed for their study}. \dc{This involves} several technical and scientific challenges including the understanding of the local deformation mechanisms that lead to a global compaction law. Such a law is expected to include the effect of variability in the particles' rheology.
Concerning the shear process of squishy systems, the main challenge consists in getting insights into the local rearrangement mechanisms, generalizing the T1-event concept. This should work as a building block for the description of the deformation of amorphous squishy materials. Studying the dynamics of this phenomenon and the spatial distribution of these events will also constitute a major step forward \dc{in understanding} the plastic processes in these systems.

These challenges cannot be overcome without tackling serious numerical and experimental difficulties. The first one consists in finding an experimental method to measure the evolution of dynamically loaded 2D and 3D systems down \dc{to} the sub-granular scale.
The numerical counterpart of this task is equally challenging since simulation methods can currently mainly deal with the quasistatic compression of a limited number of grains. 
The complexity of the simulations can rapidly increase as soon as shearing tests are considered, or realistic and strongly deformed particles geometries ought to be implemented. 

\jb{Setting up} a new framework to analyse the behavior of \textit{squishy} particle systems deep in the jammed state is an open and complex task yet to be developed in granular matter research. 
Some aspects concerning the experimental and numerical study of squishy matter are currently maturing, but major challenges are still ahead. 
Fortunately, these systems are so ubiquitous that scientists brave enough to take up these challenges can count on the support of a broad and diverse community gathering physicists, geophysicists, biologists and manufacturers to name a few.

\bibliographystyle{unsrt}
\bibliography{biblio.bib}

\end{document}